\documentclass[preprint,prc,showpacs,preprintnumbers,
               superscriptaddress,amsmath,amssymb,floatfix]{revtex4}
\usepackage{graphicx,latexsym,amssymb}
\usepackage{multirow,amsmath}

\newcommand{\ben}{\begin{enumerate}}
\newcommand{\een}{\end{enumerate}}
\newcommand{\beq}{\begin{equation}}
\newcommand{\eeq}{\end{equation}}
\newcommand{\beqn}{\begin{eqnarray}}
\newcommand{\eeqn}{\end{eqnarray}}

\newcommand{\beqd}{\begin{eqnarray*}}
\newcommand{\eeqd}{\end{eqnarray*}}
\newcommand{\bea}{\begin{array}}
\newcommand{\eea}{\end{array}}
\newcommand{\bcen}{\begin{center}}
\newcommand{\ecen}{\end{center}}
\newcommand{\btab}{\begin{tabular}}
\newcommand{\etab}{\end{tabular}}
\newcommand{\bsub}{\begin{subequations}}
\newcommand{\esub}{\end{subequations}}

\newcommand{\bfv}{\mbox{\boldmath$V$}}
\newcommand{\br}{\mbox{\boldmath$r$}}
\newcommand{\bfa}{\mbox{\boldmath$A$}}

\newcommand{\bome}{\mbox{\boldmath$\omega$}}

\newcommand{\brho}{\mbox{\boldmath$\rho$}}

\begin{document}

\title{Description of $^{178}$Hf$^{m2}$ in the
constrained relativistic mean field theory
\footnote{Supported by the National Natural Science Foundation of China under Grant Nos 10605004 and
10705004,
the Natural Science Foundation of He'nan Educational Committee under Grant No 200614003, and
the Young Backbone Teacher Support Program of He'nan Polytechnic University }}

\author{ZHANG Wei\footnote{Email: zw76@pku.org.cn}}
\affiliation{School of Physics, and State Key Laboratory of Nuclear Physics and Technology,
Peking University, Beijing 100871}
\affiliation{School of Electrical Engineering and Automation,
He'nan Polytechnic University, Jiaozuo 454003}
\author{PENG Jing}
\affiliation{Department of Physics, Beijing Normal University, Beijing 100875}
\author{ZHANG Shuang-Quan}
\affiliation{School of Physics, and State Key Laboratory of Nuclear Physics and Technology,
Peking University, Beijing 100871}

\date{\today}

\begin{abstract}
The properties of the ground state of $^{178}$Hf and the isomeric
state $^{178}$Hf$^{m2}$ are studied
within the adiabatic and diabatic constrained relativistic mean field (RMF) approaches.
The RMF calculations reproduce well the binding energy and
the deformation for the ground state of $^{178}$Hf.
Using the ground state single-particle eigenvalues obtained in the present calculation,
the lowest excitation configuration with $K^\pi=16^+$ is found to be
$\nu(7/2^-[514])^{-1}(9/2^+[624])^{1}$
$\pi(7/2^+[404])^{-1}(9/2^-[514])^{1}$.
Its excitation energy calculated by the RMF theory with time-odd fields taken into account is equal to 2.801 MeV,
i.e., close to the $^{178}$Hf$^{m2}$ experimental excitation energy 2.446 MeV.
The self-consistent procedure accounting for the time-odd component of the meson fields is
the most important aspect of the present calculation.

\end{abstract}

\pacs{21.10.-k, 21.60.-n, 27.70.+q}

\maketitle


The relativistic mean-field (RMF) theory is one of the most
successful microscopic models in nuclear physics.~\textsuperscript{\cite{Serot1986,Reinhard1989,Ring1996}}
From the very beginning, it incorporates the important relativistic effects,
and it has achieved success in describing many nuclear phenomena related to stable
nuclei \textsuperscript{\cite{Reinhard1989,Ring1996}}, exotic nuclei
\textsuperscript{\cite{Meng1996,Meng1998}} as well as supernova and
neutron stars \textsuperscript{\cite{Glendenning2000}}. The RMF
theory provides a new explanation for the identical bands in
superdeformed nuclei\textsuperscript{\cite{Konig1993}}  and for the
neutron halo in heavy nuclei \textsuperscript{\cite{Meng1996}}, it predicts giant neutron halos,
a new phenomenon in heavy nuclei close to the neutron
drip line \textsuperscript{\cite{Meng1998,Meng2002}}, it
naturally generates the spin-orbit potential, explains the origin of the pseudospin
symmetry as a relativistic symmetry
\textsuperscript{\cite{Ginocchio1997,Meng1998-2,Meng1999}}, and spin
symmetry in the anti-nucleon spectrum
\textsuperscript{\cite{Zhou2003}}, and also describes well the magnetic
rotation \textsuperscript{\cite{Madokoro2000}}, the collective
multipole excitations \textsuperscript{\cite{Ma2002}} as well as the
properties of hypernuclei\textsuperscript{\cite{Mares1994}}, etc.
Lately, the ground state properties of about 7000 nuclei have been
calculated in the RMF+BCS model and good agreements with existing
experimental data were obtained \textsuperscript{\cite{Geng2005}}.
Recent and more complete reviews of the applications of the RMF
model, particularly, those to exotic nuclei, can be found in
Refs.~\cite{Vretenar2005,Meng2006} .

Recently the 31-yr isomer of $^{178}$Hf (also called
$^{178}$Hf$^{m2}$, $K^\pi$ = $16^+$, $E_x$=2.446MeV) has attracted
extensive attention \textsuperscript {\cite{Hayes2002,Cline2005,Hayes2007,Sun2004,Tuya2006}}
for its potential to be a good medium of energy storage
\textsuperscript{\cite {Collins1999}}.
The long half-life of $^{178}$Hf$^{m2}$ is connected with
the strong inhibition of spontaneous electromagnetic transitions
restricted by the $K$ selection rule,
and it supports the point of view that this high-$K$ state has an axially symmetric intrinsic shape~\textsuperscript{\cite{Sun2004}}.
The configuration originally suggested for the isomer
$^{178}$Hf$^{m2}$ is $\nu^2(7/2^-[514])(9/2^+[624])$$\pi^2(7/2^+[404])(9/2^-[514])$~\textsuperscript{\cite {Helmer1968}} ,
which was further supported by
the view of alignment and by the $g$-factor of the corresponding rotation bands.
~\textsuperscript{\cite{Mullins1997}}

In this Letter, we will study the properties of the ground state of
$^{178}$Hf and also investigate the possible configuration of
$^{178}$Hf$^{m2}$ within the self-consistent axially symmetric RMF
theory. From the point of views of the adiabatic constrained calculation, nuclear ground
state presents the global minimum of the potential energy surface (PES).
To study the excited states, like isomers, the diabatic
(configuration-fixed) constrained approach can be applied as an effective
method.~\textsuperscript{\cite {Lv2007}} Within the adiabatic constrained
approach, nucleons always occupy the lowest levels, while in the
diabatic constrained approach, the configuration is kept fixed by
the so-called concept of ``parallel transport".
In the present paper, both the adiabatic and diabatic constrained RMF approaches are applied in
our investigation.


The basic ansatz of the RMF theory is a Lagrangian density where
nucleons are described as Dirac particles which interact via the
exchange of various mesons and the photon. The mesons considered
are the isoscalar-scalar $\sigma$, the isoscalar-vector $\omega$
and the isovector-vector $\rho$. The effective Lagrangian density reads ~\textsuperscript{\cite{Serot1986}}

\beqn \nonumber
 \cal L&=&\bar\psi \left[
i\gamma^\mu\partial_\mu-M-g_\sigma\sigma-g_\omega\gamma^\mu\omega_\mu
-g_\rho\gamma^\mu \vec\tau \cdot \vec\rho_\mu - e\gamma^\mu\dfrac{1-\tau_3}{2}A_\mu \right] \psi\\
\nonumber
&&+\dfrac{1}{2}\partial^\mu\sigma\partial_\mu\sigma-\dfrac{1}{2} m_\sigma^2\sigma^2
-\dfrac{1}{3}g_2\sigma^3-\dfrac{1}{4}g_3\sigma^4\\
\nonumber
&&-\dfrac{1}{4}\Omega^{\mu\nu}\Omega_{\mu\nu}+\dfrac{1}{2}m_\omega^2\omega^\mu\omega_\mu
+\dfrac{1}{4}c_3 (\omega^\mu\omega_\mu)^2\\
\nonumber
&&-\dfrac{1}{4}\vec R^{\mu\nu}\cdot\vec R_{\mu\nu}+\dfrac{1}{2}m_\rho^2\vec\rho^\mu\cdot\vec\rho_\mu\\
&&-\dfrac{1}{4}F^{\mu\nu}F_{\mu\nu}
\label{Lagrangian}
 \eeqn
in which the field tensors for the vector mesons and the photon
are, respectively, defined as
\begin{eqnarray}
\left\{
\begin{array}{lll}
   \Omega_{\mu\nu}   &=& \partial_\mu\omega_\nu-\partial_\nu\omega_\mu, \\
   {\vec R}_{\mu\nu} &=& \partial_\mu{\vec \rho}_\nu
                        -\partial_\nu{\vec \rho}_\mu, \\
   F_{\mu\nu}        &=& \partial_\mu A_\nu-\partial_\nu  A_\mu.
\end{array}   \right.
\label{tensors}
\end{eqnarray}

From the Lagrangian, the equation of motion for the nucleon is
\beq
    \{ \mbox{\boldmath$\alpha$}\cdot[-i\mathbf{\nabla}-\bfv(\br)]+V_0(\br)
     +  \beta [ M+S(\br) ] \} \psi_{i}
     = \varepsilon_i\psi_i,
\eeq
with the attractive scalar potential $S(\br)=g_{\omega}\omega(\br)$,
the usual repulsive vector potential
$V_0(\br)= g_{\omega} \omega_0(\br) + g_{\rho}\tau_{3} \rho_0 (\br) + e
\dfrac{1-\tau_{3}}{2} A_0(\br)$
, and the nuclear magnetic potential
$\bfv(\br)= g_{\omega} \bome(\br) + g_{\rho}\tau_{3} \brho (\br) + e
\dfrac{1-\tau_{3}}{2} \bfa(\br)$.
The Klein-Gordon equations for the mesons and electromagnetic fields are
 \beq
  (-\nabla^2+m^2_\zeta)\zeta(\br)=S_\zeta(\br),
 \eeq
where $S_\zeta(\br)$ is the source term and all other notations are the same as in Ref.
~\cite{Meng2006}.

In the RMF approaches which are widely used, only the time-even fields are essential for
the physical observables, since the time-odd components of vector fields
do not exist because of the time reversal symmetry
for the ground state of an even-even nucleus.
For an odd-$A$ or odd-odd nucleus, the unpaired valence
nucleon will give non-vanishing contribution to the nuclear
current which provides the time-odd component of vector fields, i.e., the nuclear
magnetic potential.
It is found that the nuclear magnetic potential
has small influence on the root-mean-square radii and quadrupole
moments while it plays an important role in the single-particle
properties and magnetic moments in odd-$A$ or odd-odd nuclei.\textsuperscript{\cite
{Warrier1994,Yao2006}}
One should keep in mind that for the excited states in the even-even nuclei,
there may also exist the unpaired nucleons,
which result in non-vanishing time-odd field.
Therefore the time-odd fields should also be treated carefully
for some isomeric states in the even-even nuclei,
as in the odd-$A$ or odd-odd nuclei.
In the calculation without current, the  nuclear
magnetic potential $\bfv(\br)$ will be neglected.

For the adiabatic constrained approach, the binding energy at a certain deformation is obtained
by constraining the mass quadruple moment $\langle \hat{Q_2}\rangle $
to a given value $\mu$, i.e.
\beq
  \langle H'\rangle ~=~\langle H\rangle   +   \dfrac{1}{2}C(\langle \hat{Q_2}\rangle -\mu)^2.
\eeq
where $C$ is the curvature constant parameter, and $\mu$ is
the given quadrupole moment. The expectation value of $\hat{Q_2}$ is
 $\langle \hat{Q_2}\rangle =\langle \hat{Q_2}\rangle _n+\langle \hat{Q_2}\rangle _p$,
where
 $\langle\hat{Q_2}\rangle _{n,p}= \langle 2 r^2 P_2(\cos\theta)\rangle _{n,p}$.
The deformation parameter $\beta_2$ is related to $\langle \hat{Q_2}\rangle $ by
$\langle \hat{Q_2}\rangle  = \dfrac{3}{\sqrt{5\pi}} Ar^2\beta_2$ , $r = R_0 A^{1/3}$
($R_0=1.2$ fm) and $A$ is the mass number. By varying
$\mu$, the binding energy at different deformations can be
obtained \textsuperscript {\cite{Ring1980}}.

For the adiabatic constrained approach, the occupied levels are
determined by the so-called ``parallel transport"~\textsuperscript
{\cite{Lv2007}}, i.e.,
 \beq
\langle\psi_i(q)|\psi_j(q+\Delta q)\rangle |_{\Delta q \rightarrow
0} \approx \delta _{ij},
 \eeq
where $i$ and $j$ enumerate all the single-particle levels of two
adjacent configurations. In such a way, the original configuration
at $q$ can be traced and the corresponding PES can be obtained as a
function of the deformation~\textsuperscript{\cite{Lv2007}}.
In principle, if $\Delta q$ is small
enough, the configurations at $q$ and at $q+\Delta q$ should be the
same. In the calculation, the two-step procedure is adopted: first,
the wave functions and the configuration at the initial $q$  are
recorded. Second, the wave functions $|\psi_i(q)\rangle$ are mapped
to $|\psi_j(q+\Delta q)\rangle$ one by one by searching the largest
overlap in $|\psi_j(q+\Delta q)\rangle$ with the same quantum number
$\Omega^{\pi}$. The configuration is transferred by copying the
occupation number from $|\psi_i(q)\rangle$ to the mapped
$|\psi_j(q+\Delta q)\rangle$. The wave functions and the
configuration at this $q+\Delta q$ are also recorded. The second
step is repeated until enough points on the diabatic PES are
obtained.

The constrained RMF calculations are carried out with parameter set PK1 \textsuperscript
{\cite{Long2004}}.
The full $N$ = 20 deformed harmonic-oscillator shells
for fermions and bosons are taken into account as the basis.
This basis is large enough to produce a converged binding energy at certain deformation.

\begin{figure}[t]
\includegraphics[scale=0.45]{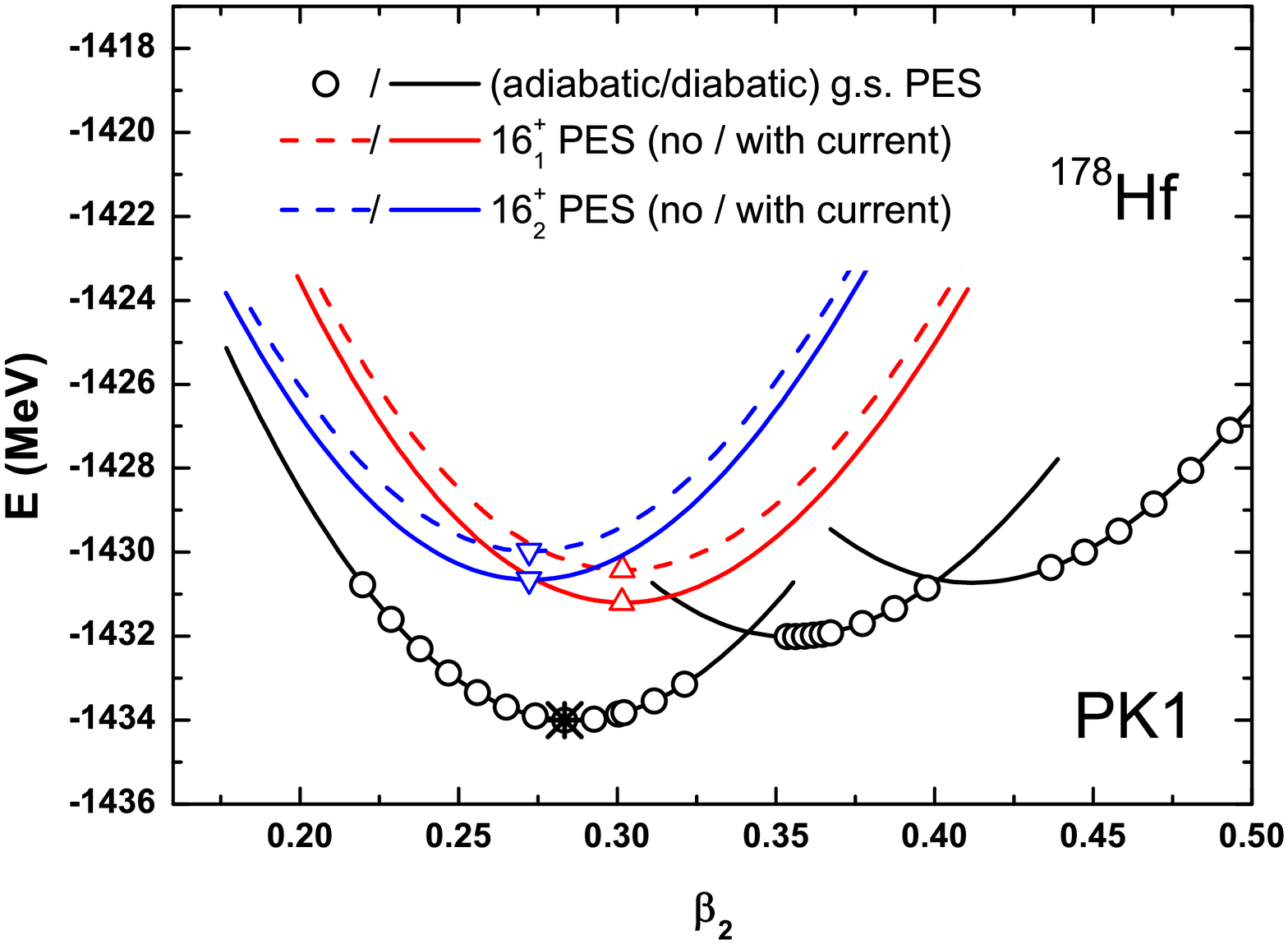}

\caption{(color online) Potential energy surfaces (PES) of $^{178}$Hf from
adiabatic (open circles)  and diabatic (solid curves) RMF calculations,
as well as the PES of the first two
$16^+$ states $16^+_1$ (red) and $16^+_2$ (blue) from the
adiabatic constrained RMF calculations  with (solid curves) and without
(dashed curves) current. The RMF parameter set PK1 is adopted.
The ground state as well as the local minima $16^+_1$ ($16^+_2$) are
denoted as a black asterisk and red (blue) triangles, respectively.
}
   \label{fig:PES}
\end{figure}

The PES of $^{178}$Hf obtained in
adiabatic (open circles) and diabatic (lines) constrained
RMF calculations are plotted in Fig.~\ref{fig:PES}. The calculated
energy $E$=-1434.0 MeV and the deformation $\beta_2$=0.283 of the
ground state (denoted as a black asterisk in Fig. \ref{fig:PES}) are in
good agreement with the experimental energy -1432.8 MeV \textsuperscript {\cite{Audi2003}}and
deformation 0.280\textsuperscript {\cite{Raman2001}}. In this figure, the adiabatic PES can be
decomposed into three regions by the discontinuity, i.e.,
$\beta_2$=0.22 $\sim$ 0.32 (region 1), $\beta_2$ = 0.35 $\sim$ 0.40
(region 2), $\beta$ = 0.43 $\sim$ 0.50 (region 3). It is known that the discontinuity
originates from the change of the configurations, and the
configuration for all points in one region is the same. This is
confirmed by the diabatic constrained calculation in which the
configuration is kept fixed during the constraint procedure. It can be seen in
Fig. \ref{fig:PES} that for every region the diabatic calculation (solid curves)
coincides with the adiabatic calculation (open circles) and extends
the region much wider. The crossover of the solid curves announces
the change of the configurations. For example, from region 1 to
region 2, the configuration changes from the ground state to $\pi
(7/2^+[404])^{-2}(1/2^-[541])^{2}$, where microscopically the
proton level $7/2^+[404]$ below the Fermi surface in region 1
becomes unoccupied while the proton level $1/2^-[541]$ above the
Fermi surface in region 1 becomes occupied. Since a pair of protons
change the levels at the same time, the $K$-values remain zero for
region 2. Similarly, from region 2 to region 3, the configuration
changes from $\pi (7/2^+[404])^{-2}(1/2^-[541])^{2}$ to $\nu
(5/2^-[512])^{-2}(1/2^+[660])^{2} \pi
(7/2^+[404])^{-2}(1/2^-[541])^{2}$ and the $K$-values also remain zero
for region 3.

Based on the single-particle spectra of the ground state,
one can construct excited states with high $K$-values.
In a deformed, axially symmetric nucleus, a high-$K$ state is made
by summing the contributions from several unpaired quasiparticles.
To form low-lying high-$K$ states, several high-$\Omega$ single-particle
(both neutron and proton) levels lying close
to the Fermi surface are necessary.
The well-deformed nuclei with $A \approx 180$, including $^{178}$Hf, satisfy this requirement very well.
With the restriction of the total $\Omega$ and parity as $16^+$,
the candidate configurations of $^{178}$Hf$^{m2}$ will be constructed.

\begin{figure}[t]
\includegraphics[scale=0.45]{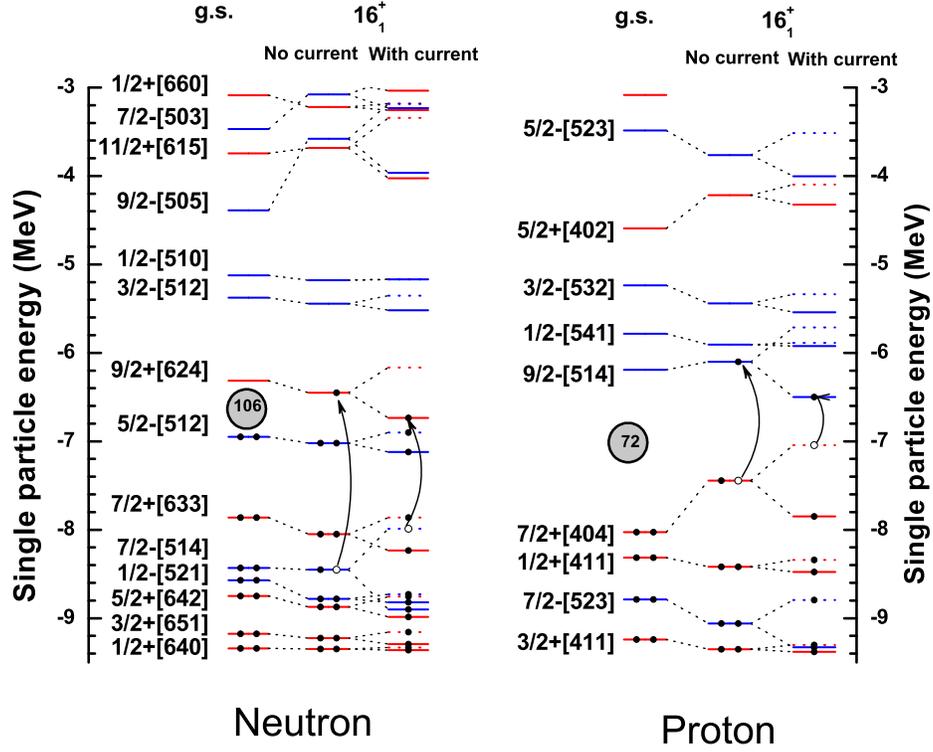}

\caption{(color online) Neutron (left panel) and proton (right panel)
single-particle levels of the ground states (1st
column), state $16^+_1$ of $^{178}$Hf obtained from the
constrained RMF approach without (2nd column) and with (3rd column) current. Each level is
labeled by the Nilsson notation $\Omega\pi[N n_z m_l]$ of its main component.
The Fermi surfaces are marked by neutron and proton numbers in the grey
circles. The red (blue) lines
denote the parity $+$ ($-$), and the dashed (solid)
lines denote the positive (negative) signs of $\Omega$, respectively.
The configuration difference is illustrated by filled
circles, open circles and arrows. }

   \label{fig:Level}
\end{figure}

For the ground state of $^{178}$Hf, the neutron (proton) single-particle levels
close to the Fermi surface are shown in the first column of the left (right) panel in Fig.~\ref{fig:Level}.
Each level is labeled by the Nilsson notation $\Omega^\pi [N n_z m_l ]$
of its main component.
It can be seen that in Fig.
\ref{fig:Level}, the energy required to excite one neutron or one proton is not
less than 0.63 or 1.84 MeV. As the experimental
excitation energy of $^{178}$Hf$^{m2}$ equals 2.446 MeV, it is sufficient
to consider one- or two-neutron excitations, together with one- or two-proton excitations.
For two-neutron (two-proton) excitations, the following cases are
considered: in the first case, a pair of particles below the Fermi surface are
excited to two different levels above the Fermi
surface; in the other case, two particles occupying different
levels below the Fermi surface are excited to form a
new pair. Those cases which involve four or more single-particle levels are
not included for simplicity. All possible configurations are
constructed by restricting the $K$ (total $\Omega$) value to 16 and
the nuclear parity to $+$. Thus the configurations with the lowest excitation
energies are obtained and labeled by $16^+_1$ ,$16^+_2$, $16^+_3$,
etc.

The detailed configurations of the five lowest $K^\pi=16^+$ states of
$^{178}$Hf are listed in column 2 of Table \ref{Tab:tab}.
In this table, the first 4 states are one-neutron plus one-proton
excitation, while the last one is a pair of neutrons excited to
high-$\Omega$ levels combined with the one-proton excitation.
The configuration of $16^+_1$ is $\nu(7/2^-[514])^{-1}(9/2^+[624])^{1}$
$\pi(7/2^+[404])^{-1}(9/2^-[514])^{1}$ with respect to the ground state, i.e.,
one formerly paired neutron in the level $7/2^-[514]$ becomes unpaired
and excited to the level $9/2^+[624]$, and another formerly
paired proton in level $7/2^+[404]$ is excited to the level $9/2^-[514]$.
This configuration is consistent with the former assignment of $^{178}$Hf$^{m2}$
\textsuperscript{\cite{Helmer1968,Mullins1997}}.
Note that the excitation energy  of $16^+_1$ given by the sum of single
particle excitations is 3.954 MeV which is
about 1.5 MeV higher than the experimental value 2.446 MeV.

In order to obtain the self-consistent excitation energy for
state $16^+_1$ microscopically, the diabatic constrained RMF calculations are carried
out with the respective configuration information.
As discussed before, the time-odd fields \textsuperscript{\cite{Yao2006}} caused by the unpaired nucleons
should be taken into account carefully for the high-$K$ state.
The time-even calculation is also done for comparison. The PES of state
$16^+_1$ with (without) current is plotted as a solid (dashed) red
curve in Fig.~\ref{fig:PES}, together with the PES of state $16^+_2$.
The local minima of $16^+_1$ ($16^+_2$) PES are denoted as up (down)
triangles in Fig.~\ref{fig:PES}.
For state $16^+_1$, the excitation energy according to the calculations
without and with current is, respectively 3.579 MeV and 2.801 MeV, which
clearly shows that both the self-consistent calculation and the consideration of the time-odd
fields are crucial effects to obtain the reasonable excitation energy.
The excitation energy of $16^+_1$ ($E_x=2.801$ MeV) is close to the experimental
excitation energy ($E_x=2.446$ MeV) of $^{178}$Hf$^{m2}$.
The deformation ($\beta_2=0.30$) obtained for this isomer is similar to
that of the ground state.
In Table \ref{Tab:tab}, all the excitation energies from calculations without and with current
as well as the deformation calculated for the states $16^+_1$ to $16^+_5$ are also given.
For all five states, the excitation energies calculated with
current are 0.6 $\sim$0.8MeV smaller than those without current.
By the way, the same approaches are also applied to the isomeric state $^{178}$Hf$^{m}$ (experimentally $K^\pi$=$8^-$,$E_x$=1.147MeV).
The configuration obtained for $8^-_1$ is $\pi(7/2^+[404])^{-1}(9/2^-[514])^{1}$, and the excitation energy
without and with current is, respectively 1.552 MeV and 1.315 MeV.

In Fig.~\ref{fig:Level}, the neutron and proton single-particle levels
of state $16^+_1$ from the RMF without and with current
have been plotted in columns 2, 3 in both panels.
In the time-odd calculation each single-particle level splits into two levels
due to the breaking of the time-reversal symmetry,
the level with positive $\Omega$ being
energetically favored. Such splittings will
change straightforward the energy gap between two single-particle levels.
In particular, for one-neutron and one-proton excitation of the state $16^+_1$ discussed here,
the neutron energy gap between $\nu(9/2^+[624])$ and
$\nu(7/2^-[514])$ in the calculation without current
decreases from 2.00 MeV to 1.25 MeV , which is the
gap between $\nu(9/2^+[624],+\Omega)$ and
$\nu(7/2^-[514],-\Omega)$ in the calculation with current.
Correspondingly the energy gap for the proton excitation concerned
decreases from  1.34 MeV to 0.54 MeV.
As a result, the calculation with current decreases the excitation energy considerably.

In summary, the properties of the ground state of $^{178}$Hf and
the isomeric state $^{178}$Hf$^{m2}$ are investigated
by the adiabatic and diabatic constrained RMF approaches.
The constrained RMF theory reproduces well the binding energy and
deformation for the ground state of $^{178}$Hf.
With the single-particle levels of the ground state obtained in the adiabatic constrained RMF theory,
by restricting $K^\pi=16^+$,
the configuration with the lowest excitation energy is found to be
$\nu(7/2^-[514])^{-1}(9/2^+[624])^{1}$
$\pi(7/2^+[404])^{-1}(9/2^-[514])^{1}$,
which is consistent with the former configuration assignment.
The excitation energy based on the single-particle spectra is 3.954 MeV, which is
much higher than the experimental excitation energy 2.446 MeV of $^{178}$Hf$^{m2}$.
By applying the self-consistent time-even and time-odd RMF calculation,
the excitation energy of this configuration is decreased to 3.579 MeV and 2.801 MeV, respectively.
Therefore both the self-consistency and the consideration of current are
important factors for studying nuclear isomers.

The authors gratefully acknowledge Professor Jie Meng, Professor L. N. Savushkin,
and Dr. Jiangming Yao for their helpful suggestions and discussions.

\renewcommand{\baselinestretch}{1}
\small \normalsize

\clearpage

\begin{table}
\caption{The configurations, excitation energies as well as the deformations obtained
from the adiabatic and diabatic constrained RMF approaches for
the first five $K^\pi=16^+$ states of
$^{178}$Hf. The excitation energies (in MeV) include
the sum of single-particle excitations
$\displaystyle\sum(\varepsilon_j-\varepsilon_i)$, and
the excitation energies $E_x$ calculated without (with) current.
}
\small
\begin{center}
\begin{tabular}{ccccccc}
\hline

\multirow{2}{0.2cm}{State} & \multirow{2}{2cm}{Configuration}
 & \multicolumn{2}{c}{\multirow{2}{2cm}{$\displaystyle\sum(\varepsilon_j-\varepsilon_i)$}}&$E_x$ & $E_x$ &$\beta_2$\\
 &  & & &(no current)& (with current)& (with current)\\

\hline \multirow{2}{0.3cm}  {$16^+_1$}&
$\nu(7/2^-[514])^{-1}(9/2^+[624])^{1}$& 2.119&\multirow{2}{0.3cm}
{3.954}
&\multirow{2}{0.3cm}  {3.579}&\multirow{2}{0.3cm}  {2.801} &\multirow{2}{0.3cm}  {0.302} \\
&
$\pi(7/2^+[404])^{-1}(9/2^-[514])^{1}$&     1.835&          &&&\\
\hline \multirow{2}{0.3cm}  {$16^+_2$}&
$\nu(7/2^+[633])^{-1}(9/2^+[624])^{1}$& 1.549&\multirow{2}{0.3cm}
{4.146}
&\multirow{2}{0.3cm}  {4.025}&\multirow{2}{0.3cm}  {3.342} &\multirow{2}{0.3cm}  {0.272} \\
&
$\pi(7/2^-[523])^{-1}(9/2^-[514])^{1}$&     2.597&          &&&\\
\hline \multirow{2}{0.3cm}  {$16^+_3$}&
$\nu(5/2^-[512])^{-1}(11/2^+[615])^{1}$& 3.204&\multirow{2}{0.3cm}{5.039}
&\multirow{2}{0.3cm}  {4.749}&\multirow{2}{0.3cm}  {4.129}&\multirow{2}{0.3cm}  {0.291} \\
&
$\pi(7/2^+[404])^{-1}(9/2^-[514])^{1}$&     1.835&          &&&\\
\hline \multirow{2}{0.3cm}  {$16^+_4$}&
$\nu(7/2^+[633])^{-1}(9/2^-[505])^{1}$&     3.473&\multirow{2}{0.3cm}{5.308}
&\multirow{2}{0.3cm}  {5.380}&\multirow{2}{0.3cm}  {4.601}&\multirow{2}{0.3cm}  {0.273} \\
&
$\pi(7/2^+[404])^{-1}(9/2^-[514])^{1}$&     1.835&          &&&\\
\hline \multirow{2}{0.3cm}  {$16^+_5$}&
$\nu(5/2^-[512])^{-2}(9/2^+[624])^{1}(7/2^-[503])^{1}$&     4.113&\multirow{2}{0.3cm}  {5.948}
&\multirow{2}{0.3cm}  {5.879}&\multirow{2}{0.3cm}  {5.302}&\multirow{2}{0.3cm}  {0.290} \\
&
$\pi(7/2^+[404])^{-1}(9/2^-[514])^{1}$&     1.835&          &&&\\
\hline
\normalsize
\end{tabular}
\end{center}
   \label{Tab:tab}
\end{table}

\end{document}